\documentclass[aps,twocolumn,showpacs,letterpaper]{revtex4-1}
\usepackage{amsmath,amssymb}
\usepackage{hyperref}
\usepackage{graphicx}
\usepackage{dcolumn}
\usepackage{listings}
\usepackage{color}

\hypersetup{pdfnewwindow=true, colorlinks=true, linkcolor=blue, anchorcolor=blue,
  citecolor=blue, filecolor=blue, menucolor=blue, urlcolor=blue}

\begin{document}

\def\EF{$E_\textrm{F}$}
\def\cred{\color{red}}
\def\cblue{\color{blue}}
\definecolor{dkgreen}{rgb}{0.31,0.49,0.16}
\def\cgreen{\color{dkgreen}}

\title{Material design of indium based compounds:
possible candidates for charge, valence, and bond disproportionation and superconductivity}

\author{Chang-Jong Kang$^1$}
\email[]{ck620@physics.rutgers.edu}
\author{Gabriel Kotliar$^{1,2}$}
\affiliation{
$^1$Department of Physics and Astronomy, Rutgers University,
Piscataway, New Jersey 08854, USA \\
$^2$Condensed Matter Physics and Materials Science Department,
Brookhaven National Laboratory, Upton, New York 11973, USA
}
\date{\today}

\begin{abstract}
We design and investigate the physical properties of new indium compounds
AInX$_{3}$ (A = alkali metals, X = F or Cl).
We find nine new In based materials in their ground state
that are thermodynamically stable
but are not reported in ICSD (Inorganic Crystal Structure Database).
We also discuss several metastable structures.
This new series of materials display multiple valences,
charge and bond disproportionation, and dimerization.
The most common valence of In is  3+.
We also find two rare alternatives, one has In$^{2+}$ with In-In dimerization
and the other shows valence disproportionation to In$^{1+}$ and In$^{3+}$
with bond disproportionation.
We study  the  possibility of superconductivity in
these new In compounds and find that
CsInF$_{3}$ has a transition temperature about 24 K
with sufficient hole doping and pressure.
\end{abstract}


\maketitle

\section{Introduction}
\label{sec:intro}

Progress in theory and  algorithms,  coupled with increased computational power is now  accelerating the discovery of  new materials with useful physical properties.
A notable recent success is the theory led discovery of high temperature superconductivity of hydrogen sulfide under pressure \cite{Duan14,Drozdov14,Drozdov15,Errea15,Ge16},
which has set the record of the highest critical temperature reached so far.
Other examples of recent predictions are existence of a metallic layer in a new family (112) of iron based superconductors \cite{Shim09,Katayama13,Kang17},
the prediction and synthesis of missing half-Heusler compounds \cite{Zhang12,Yan15,Gautier15}, the prediction of new high-pressure phase materials such as FeO$_{2}$ \cite{Hu16}, calcium carbides \cite{Li14}, and Na$_{2}$He \cite{Dong17}, which were also confirmed experimentally. For reviews see Refs.~\cite{Curtarolo13,Saal13,Butler16,Jain16,Jain16-apl,Butler18}.
Theory and computation  are thus  beginning to play a decisive role in the search for new materials.

In this paper we  apply  material design  methodology to find  new  indium halide compounds.  There are multiple motivations for this study.   One is  the intrinsic interest in finding new mixed valent compounds, a condition that is very rare in the solid state, but has been a focus of interest for many years \cite{Varma76}.
With an eye on possible applications, halide perovskites, such as CsPbI$_{3}$ \cite{Eperon15}, are remarkable photovoltaic \cite{Filip15,Korbel16} materials.
They have also been studied as analogs, as they have very similar band structures     \cite{Retuerto13,Yin13epl} of the high temperature superconductor BaBiO$_{3}$ \cite{Cava88}.
Another motivation is to make predictions in a new arena that will test a material design methodology, which assigns a likelihood that a material will form.

Indium compounds tend to form in an oxidation state of In$^{3+}$, and the valences
In$^{1+}$ and In$^{2+}$ are not typical \cite{Fedorov17,Pardoe07,Davidovich16}.
We find nine new indium fluoride/chloride compounds  where indium is in a valence 2+ or 1+, which have a very high probability to form within the framework of Adler \emph{et al.} \cite{Adler17} and are thermodynamically stable within the density functional theory PBE (Perdew-Burke-Ernzerhof) functional.
For a given composition we identify potential low-energy crystal structures for the indium fluoride/chloride compounds, as candidate structures and find the relation between the crystal structure and the valence state of indium.
These indium fluoride/chloride compounds display multiple valences, charge and bond disproportionation, and indium-indium dimerization.
The new indium fluoride/chloride compounds have too large of a band gap (of the order of  $\sim$4 eV) to be useful for photovoltaic materials, but could be useful for other applications. In particular, hole-doped CsInF$_{3}$ under pressure will exhibit superconductivity with transition temperature of 24 K.

\section{Outline}
\label{sec:outline}

In the beginning, we would like to summarize a material design workflow
introduced in Ref.~[\onlinecite{Adler17}].
The material design process is initiated by the \textbf{qualitative ideas}
including some physical idea of a model that one would like to explore or test,
ways to enhance desirable physical properties of a material, and
comparisons of a class of compounds that exhibit similar physical properties.
The initial intuitions (zeroth order step)
could be refined with simplified quantitative calculations
using a model Hamiltonian or other computational tools
like \emph{ab initio} density functional theory (DFT).
In this work, we have used DFT as a main computational tool to design a new material.
Note that the ``theoretical'' material design workflow
progresses in reverse order from experimental solid state synthesis.

The first step is the quantitative calculation of the \textbf{electronic structure}.
It explores how to go from a well defined crystal structure of a material
to the physical or chemical properties.
The second step is the \textbf{prediction of the crystal structure}
given a fixed chemical composition.
There are a number of structure prediction techniques \cite{Woodley08}
with DFT total energy calculations,
and the prediction techniques require having an accurate method
for calculating the total energy of a material at the DFT level.
The third step is testing for \textbf{thermodynamic stability}:
given the lowest energy crystal structure of the fixed composition,
check whether it is stable against
decomposition to all other compositions in the chemical system.
The third step requires the knowledge of all other known stable compositions,
their crystal structures and total energies,
which is now facilitated by materials databases,
such as Materials Project \cite{Jain13,MP}, OQMD \cite{Kirklin15,OQMD},
and AFLOWlib \cite{Setyawan11,AFLOW}.
Here, we have used the Materials Project database
for analysis of thermodynamic stability
and estimated the existence probability \cite{Adler17} based on the database.

Once the (thermodynamically) stable crystal structure is found,
the electronic structure is calculated again with more elaborated DFT methods
like hybrid functionals, the modified Becke-John (mBJ) exchange potential,
and $GW$ methods to obtain more accurate physical properties.

This paper is organized as follows:
the detail computational method of each step for material design
is provided in Sec.~\ref{sec:method}.
The computational results are presented in sequence of the material design workflow.
Section~\ref{sec:results_initial} summarizes
the initial qualitative idea (zeroth order step) and its validation (first step).
The prediction of crystal structures (second step) of the indium halide compounds
and their thermodynamic stabilities (third step)
are presented in Sec.~\ref{sec:results_crystal}
and Sec.~\ref{sec:results_stability}, respectively.
The details of electronic structures and superconducting properties (post-process)
are provided in Sec.~\ref{sec:results_electronics}.
The paper closes
with brief summary and conclusions in Sec.~\ref{sec:conclusion}.

\section{Method}
\label{sec:method}

\subsection{Prediction of crystal structure}
To obtain low-energy crystal structures of AInX$_{3}$ (A = alkali metals, X = F or Cl),
we employed the \emph{ab initio} evolutionary algorithm \cite{Oganov06}
implemented in USPEX \cite{uspex}
combined with DFT
pseudopotential code VASP \cite{VASP1,VASP2}.
The initial structures were randomly generated
according to possible space groups.
In these calculations,
the structural optimization of all the newly generated structures
were carried out by VASP
with an energy cutoff of 520 eV and the exchange-correlation
functional of generalized gradient approximation (GGA) of
Perdew-Burke-Ernzerhof (PBE) \cite{PBE}
with the projector augmented wave (PAW) method \cite{PAW1,PAW2}.

\subsection{Thermodynamic stability}

After the low-energy crystal structures were obtained,
the corresponding total energies were calculated
by using DFT pseudopotential code VASP \cite{VASP1,VASP2}.
For the Brillouin-zone integration,
an equal $k$-grid density was used for all materials
by adopting the Monkhorst-Pack sampling grid
with a reciprocal spacing of 64 $k$-points per {\AA}$^{-1}$ .
We used the empirical correction schemes
employed in Materials Project \cite{Wang06,Ong07,Jain11,Hautier12,Jain13,Adler17}
to get an accurate formation energy.
In addition, the Materials Project database was used
for analysis of thermodynamic stability
and the existence probability \cite{Adler17}.

\subsection{Electronic structure}
The all-electron full-potential linearized augmented plane-wave (FP-LAPW) method implemented in WIEN2k \cite{Wien2k} was adopted to calculate the electronic structure.
The GGA(PBE) exchange-correlation functional
was chosen to calculate the electronic structure.
To get the precise band gap,
we utilized the modified Becke-John (mBJ) exchange potential \cite{Tran09},
which is rather accurate and computationally cheaper than the $GW$ method.
The Brillounin zone integration was done with a 17$\times$17$\times$17 $k$-mesh
and the plane-wave cutoff was $R_{\text{mt}}K_{\text{max}}$ = 7.
The maximum $L$ value for the waves inside the atomic spheres, $L_{\text{max}}$ = 10,
and the largest $G$ in the charge Fourier expansion $G_{\text{max}}$ = 12
were used in the calculations.

\begin{figure*}[t]
\includegraphics[width=16.0 cm]{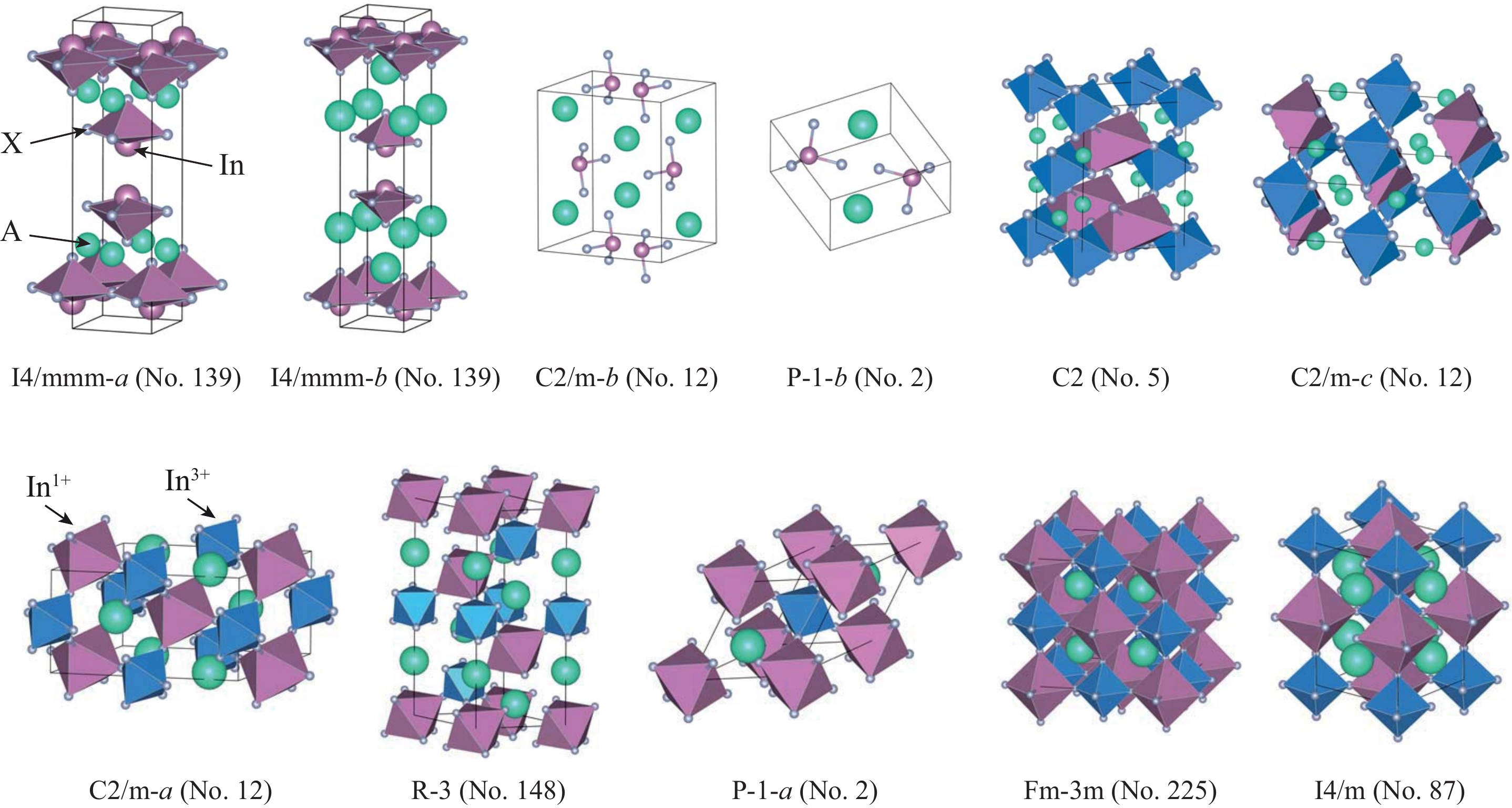}
\caption{(Color Online)
Possible candidates for low-energy crystal structures of AInX$_{3}$
(A = alkali metals, X = F or Cl)
obtained by USPEX simulations.
Space groups are also provided below for each crystal structure.
Structures listed in the second row
could be identified from structural distortions
(breathing and/or tilting of InX$_{6}$ octahedra)
in the ideal cubic perovskite structure.
For the structures,
octahedra centered at In$^{3+}$ or In$^{1+}$
are colored in navy or pink, respectively.
Letters $a$, $b$, and $c$ are attached to the end of the space group notation
if different crystal structures have the same space group.
}
\label{crystal}
\end{figure*}

\subsection{Phonon dispersion and electron-phonon coupling calculations}
We used the linear response method \cite{Baroni01}
implemented in Quantum Espresso \cite{QE} for phonon calculations.
All pseudopotentials used in the calculations were
adopted from Standard Solid State Pseudopotentials \cite{Lejaeghere16,SSSP}.
We used a 10 $\times$ 10 $\times$ 10 $k$-grid and a Gaussian smearing of 0.03 Ry
for the electronic integration.
The dynamical matrices were calculated
on a 4 $\times$ 4 $\times$ 4 phonon-momentum grid.
A 20 $\times$ 20 $\times$ 20 $k$-grid was used for the electron-phonon coupling calculations.
The standard exchange-correlation functionals
like local-density approximation (LDA)
and generalized gradient approximation (GGA)
neglect the long-range exchange interaction.
This nonlocal correlation could play a significant role
in electron-phonon interaction \cite{Yin13}.
To incorporate the long-range exchange interaction,
the Heyd-Scuseria-Ernzerhof hybrid functional (HSE06) \cite{Krukau06} was used
and the electron-phonon coupling constant was calculated
from the HSE06 deformation potential
introduced in Ref.~[\onlinecite{Yin13}].

\section{Computational results and discussion}

\subsection{Initial idea and its validation}
\label{sec:results_initial}

We conceived the idea that new indium halide compounds
have physical and chemical similarities
with BaBiO$_{3}$ \cite{Cava88}, a high temperature superconductor,
and CsPbI$_{3}$ \cite{Eperon15},
a high performance photovoltaic material.
To justify the initial idea briefly,
we assumed that new indium halide compounds have the ideal cubic perovskite structure.
To get an unknown lattice constant of the perovskite structure,
we performed the structural relaxation with VASP.
After obtaining the relaxed crystal structures for the indium halide compounds,
the electronic structures were obtained from the DFT calculations.
Comparing the electronic structures,
we could conclude that indium halide compounds
have similar physical properties to the ones exhibited in BaBiO$_{3}$ and CsPbI$_{3}$.
Therefore, indium halide compounds could be possible candidates for
high-temperature superconductors or photovoltaic materials.

\subsection{Crystal structure and total energy profile}
\label{sec:results_crystal}

\begin{figure*}[t]
\includegraphics[width=17.0 cm]{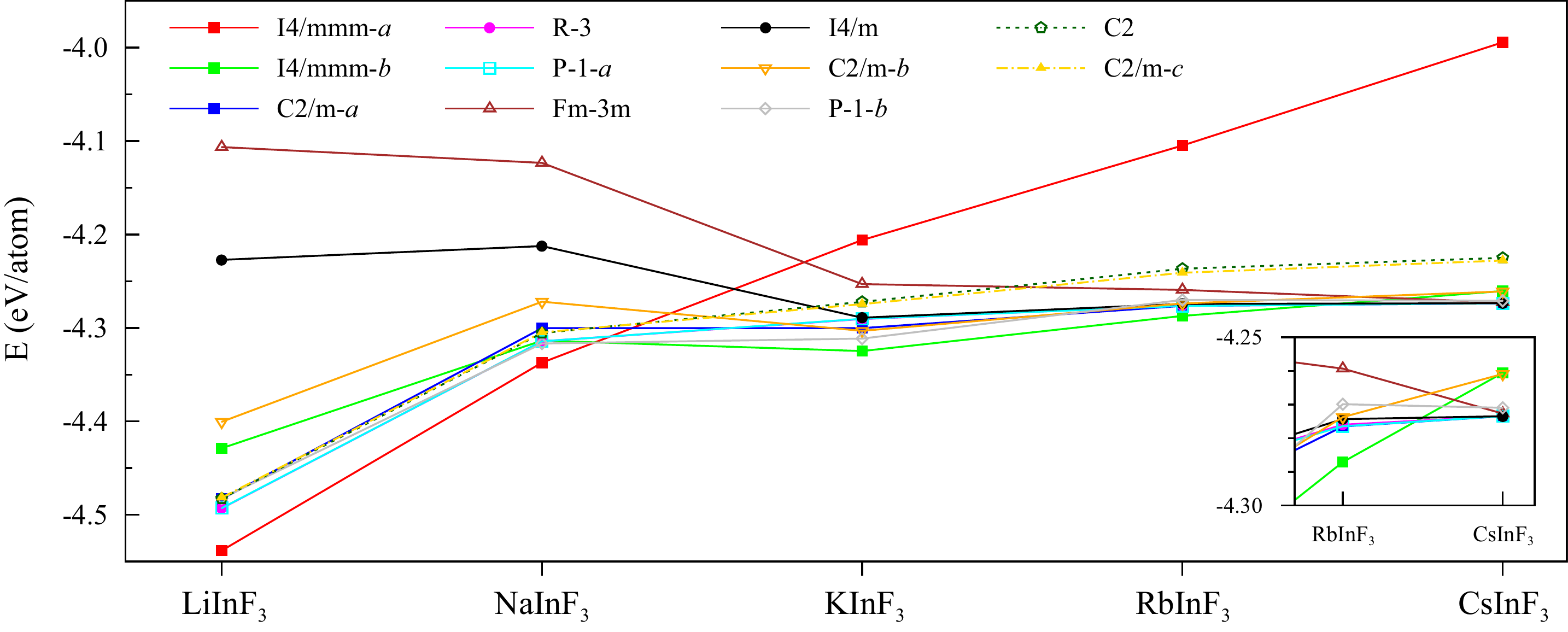}
\caption{(Color Online)
Total energies of crystal structures listed in Fig.~\ref{crystal}
for AInF$_{3}$ (A = alkali metals).
Inset is provided to zoom in
for nearly degenerate phases of RbInF$_{3}$ and CsInF$_{3}$.
The space group $I4/mmm$-$a$ is the ground state crystal structure
for both LiInF$_{3}$ and NaInF$_{3}$, and
$I4/mmm$-$b$ is for both KInF$_{3}$ and RbInF$_{3}$.
For CsInF$_{3}$, five space groups
($C2/m$-$a$, $R\bar{3}$, $P\bar{1}$-$a$, $Fm\bar{3}m$, and $I4/m$)
are candidates for the lowest energy structure
and are almost degenerate.
The energy differences among the five space groups are within 1 meV/atom.
}
\label{AInF3}
\end{figure*}

\begin{figure*}[t]
\includegraphics[width=17.0 cm]{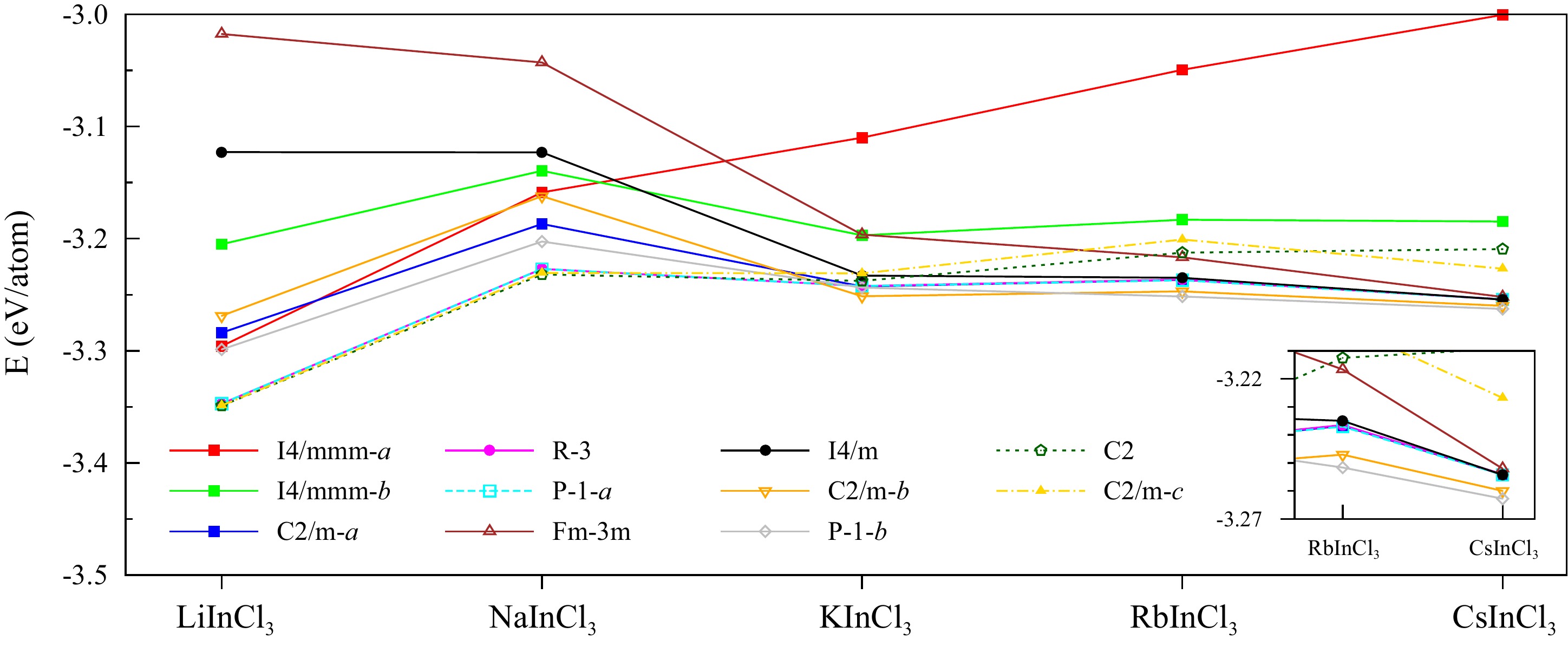}
\caption{(Color Online)
Total energies of crystal structures listed in Fig.~\ref{crystal}
for AInCl$_{3}$ (A = alkali metals).
Inset is provided to zoom in
for nearly degenerate phases of RbInCl$_{3}$ and CsInCl$_{3}$.
Four space groups, $R\bar{3}$, $P\bar{1}$-$a$, $C2$, and $C2/m$-$c$, are candidates
for the lowest energy structure of LiInCl$_{3}$ and NaInCl$_{3}$,
and energy differences among the four space groups are within 5 meV/atom.
For KInCl$_{3}$, RbInCl$_{3}$, and CsInCl$_{3}$,
two space groups, $C2/m$-$b$ and $P\bar{1}$-$b$, are candidates
for the lowest energy structure,
and energy differences between the two space groups are within 9 meV/atom.
}
\label{AInCl3}
\end{figure*}

The possible candidates for low-energy crystal structures of
AInX$_{3}$ (A = alkali metals, X = F or Cl)
obtained by USPEX are summarized in Fig.~\ref{crystal} \cite{structure-set}.

The crystal structures having the lowest energy
are tetragonal structures with space group $I4/mmm$ (No. 139)
for AInF$_{3}$ (A = Li, Na, K, and Rb) (see Fig.~\ref{AInF3})
however, Wyckoff sites for A are different:
$4d$ for both Li and Na, and $4e$ for both K and Rb.
Therefore, we have labeled the space group $I4/mmm$
having two different Wyckoff sites $4d$ and $4e$
with $I4/mmm$-$a$ and $I4/mmm$-$b$, respectively,
in order to distinguish different crystal structures with the same space group.
The coordination number of In in $I4/mmm$-AInF$_{3}$ (A = Li, Na, K, and Rb) is 5
that makes a InF$_{5}$ pyramid (Fig.~\ref{crystal}).
Alkali metals A and F strongly have 1+ and 1- valence states, respectively,
hence, In should have 2+ valence state in order for the system to be charge neutral.
The 2+ valence state for In is very unique because In usually has 3+ valence state.

For CsInF$_{3}$,
five space groups, $C2/m$-$a$, $R\bar{3}$, $P\bar{1}$-$a$, $Fm\bar{3}m$, and $I4/m$
are candidates for the lowest energy structure
and have tiny energy differences within 1 meV/atom
(which corresponds to $\sim$10 K in temperature) to each other (Fig.~\ref{AInF3}).
The details of the energy differences could be found in Table~\ref{stability-F}.
They could be identified from structural distortions
(breathing and/or tilting of InF$_{6}$ octahedra)
in the ideal cubic perovskite structure
and have different sizes of InF$_{6}$ octahedra (Fig.~\ref{crystal}).
One has short In-F bond length ($\sim$2.12 {\AA})
and it leads for In having 3+ valence state.
On the other hand, the other has long In-F bond length ($\sim$2.68 {\AA})
and it results in In 1+ valence state.
Therefore, the different In-F bond lengths
(bond disproportionation) trigger charge/valence disproportionation.
We would like to note that
the crystal structure of CsInF$_{3}$ with the space group $C2/m$-$a$
is quite similar to that of BaBiO$_{3}$,
which has also the monoclinic structure
with a space group $C2/m$ at low temperature \cite{Cox76,Thornton78,Cox79}.
For BaBiO$_{3}$, the Bi atoms occupy two distinct Wyckoff sites
with average Bi-O distances of
2.28 and 2.12 {\AA}, respectively \cite{Cox76},
which results in valence disproportionation
of Bi$^{3+}$ and Bi$^{5+}$ \cite{Orchard77,Cox79}.

The crystal structure with the space group $I4/mmm$-$a$
has the lowest energy for LiInF$_{3}$ and NaInF$_{3}$,
however, it has higher energy as the alkali metal gets heavier
as shown in Fig.~\ref{AInF3}.
The other crystal structures listed in Fig.~\ref{crystal}
have significant energy differences among them in LiInF$_{3}$
but these energy differences become smaller as the alkali metal becomes heavier.
For CsInF$_{3}$, these energy differences are within 13 meV/atom
and energy differences among five candidate structures
(with space groups $C2/m$-$a$, $R\bar{3}$, $P\bar{1}$-$a$, $Fm\bar{3}m$, and $I4/m$)
are only within 1 meV/atom.
The energy difference between the lowest and the next lowest energy
is gradually reduced as the alkali metal becomes heavier:
45, 20, 13, 11, and 0.01 meV/atom for Li, Na, K, Rb, and Cs compounds,
respectively (see Table~\ref{stability-F}).
Judging from these energy differences,
AInF$_{3}$ is likely synthesized into the space group $I4/mmm$-$a$ for both Li and Na
and $I4/mmm$-$b$ for both K and Rb compounds if they are thermodynamically stable
(their thermodynamic stabilities will be discussed later on).
However, synthesis of CsInF$_{3}$ quite depends on
the synthesis method and condition due to the quite small energy difference.
It could be one of five candidate structures
(with space groups $C2/m$-$a$, $R\bar{3}$, $P\bar{1}$-$a$, $Fm\bar{3}m$, and $I4/m$)
depending on the synthesis condition.
Or otherwise, it exists in several different polymorphs.
Note that charge-ordered thallium halide perovskites
CsTlX$_{3}$ (X = F or Cl) are successfully synthesized.
CsTlF$_{3}$ has a cubic phase ($Fm\bar{3}m$).
On the other hand,
CsTlCl$_{3}$ exists in two different polymorphs: a tetragonal phase ($I4/m$) and
a cubic phase ($Fm\bar{3}m$) \cite{Retuerto13}.

The possible candidates for low-energy crystal structures of AInCl$_{3}$
have been also investigated along with AInF$_{3}$.
Since fluorine and chlorine atoms belong to the same halogen group
in the periodic table,
AInCl$_{3}$ is expected to have close resemblances to AInF$_{3}$
structurally and electronically.
To check the possible low-energy crystal structures for AInCl$_{3}$,
we examined the total energies of the crystal structures listed in Fig.~\ref{crystal}
and the results are shown in Fig.~\ref{AInCl3}.
For both LiInCl$_{3}$ and NaInCl$_{3}$,
four space groups, $R\bar{3}$, $P\bar{1}$-$a$, $C2$, and $C2/m$-$c$,
are candidates for the lowest energy structure
and have tiny energy differences only within 5 meV/atom.
The details of the energy differences could be found in Table~\ref{stability-Cl}.
These four space groups, $R\bar{3}$, $P\bar{1}$-$a$, $C2$, and $C2/m$-$c$,
possess two different sizes of InCl$_{6}$ octahedra
(bond disproportionation, see Fig.~\ref{crystal}),
which leads to In having In$^{1+}$ and In$^{3+}$ valence states
and showing charge/valence disproportionation.
For AInCl$_{3}$ (A = K, Rb, Cs),
two space groups, $C2/m$-$b$ and $P\bar{1}$-$b$,
are candidates for the lowest energy structure.
Energy differences between the two space groups are
7.8, 4.6, and 2.8 meV/atom for K, Rb, and Cs compounds,
respectively (see Table~\ref{stability-Cl}).
These two space groups possess symmetrically equivalent In atoms in the unit cell,
hence charge/valence disproportionation is not available for these space groups
(bond disproportionation is not available as well).
The coordination number of In in both $C2/m$-$b$ and $P\bar{1}$-$b$ space groups is 3,
which is quite distinct from the others listed in Fig.~\ref{crystal}.
The space group $I4/mmm$-$a$ has higher energy as the alkali metal gets heavier
as shown in Fig.~\ref{AInCl3}.
Besides, energy differences among space groups except for $I4/mmm$-$a$
are significant for LiInCl$_{3}$
and become smaller as the alkali metal becomes heavier.
These two tendencies are also realized in AInF$_{3}$ compounds (Fig.~\ref{AInF3}).

\begin{figure*}[t]
\includegraphics[width=15.3 cm]{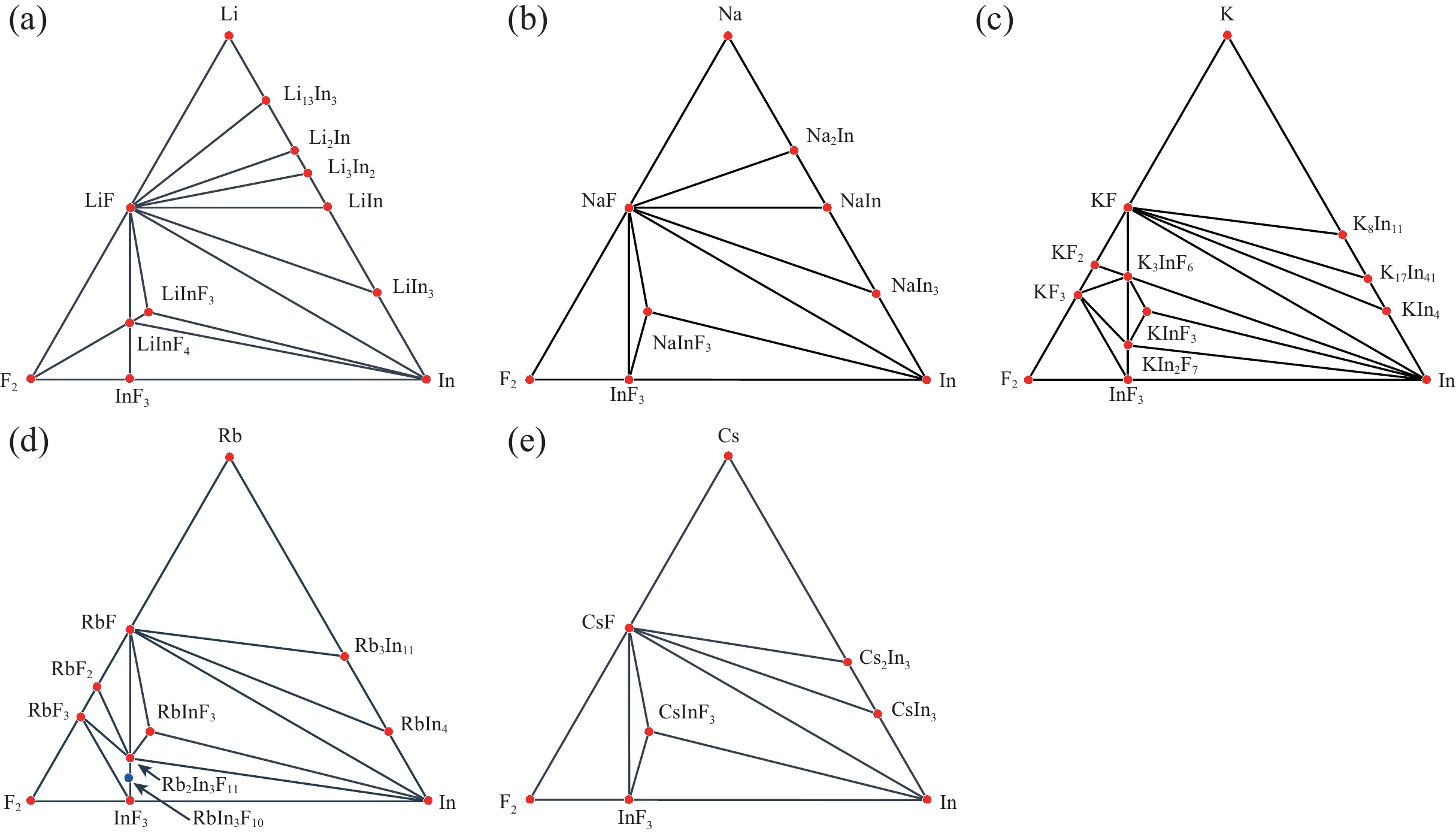}
\caption{(Color Online)
Ternary phase diagrams for (a) LiInF$_{3}$,
(b) NaInF$_{3}$, (c) KInF$_{3}$, (d) RbInF$_{3}$,
and (e) CsInF$_{3}$.
Note that RbIn$_{3}$F$_{10}$ is marginally stable
with a tiny energy above hull of 1.26 meV/atom
(existence probability of 0.43).
}
\label{ternary-F}
\end{figure*}

\begin{table}[b]
\caption{
Hull energy ($E_{\text{Hull}}$), energy difference ($\Delta E$),
and existence probability
for the low-energy crystal structure of AInF$_{3}$ (A = alkai metals).
The hull energy is obtained from the determinant reaction (see text),
the energy difference is estimated
with respect to the crystal structure having the lowest energy,
and the existence probability is calculated
by using the probabilistic model introduced in Ref.~[\onlinecite{Adler17}].
The units for $E_{\text{Hull}}$ and $\Delta E$ are meV/atom.
}
\begin{ruledtabular}
\begin{tabular}{c|c|c|c}
Space group & $E_{\text{Hull}}$ (meV) & $\Delta E$ (meV) & Prob. \\
\hline
\multicolumn{4}{c}{LiInF$_{3}$} \\
\hline
$I4/mmm$-$a$ (No. 139) & -30.362 & 0 & 0.56 \\
$R\bar{3}$ (No. 148)   & 14.860 & 45.222 & 0.39 \\
\hline
\multicolumn{4}{c}{NaInF$_{3}$} \\
\hline
$I4/mmm$-$a$ (No. 139) & -54.183 & 0 & 0.64 \\
$P\bar{1}$-$b$ (No. 2) & -33.838 & 20.345 & 0.57 \\
\hline
\multicolumn{4}{c}{KInF$_{3}$} \\
\hline
$I4/mmm$-$b$ (No. 139) & -20.501 & 0 & 0.52 \\
$P\bar{1}$-$b$ (No. 2) & -7.298 & 13.202 & 0.46 \\
\hline
\multicolumn{4}{c}{RbInF$_{3}$} \\
\hline
$I4/mmm$-$b$ (No. 139) & -49.727 & 0 & 0.62 \\
$C2/m$-$a$ (No. 12)    & -39.139 & 10.588 & 0.59 \\
$P\bar{1}$-$a$ (No. 2) & -39.109 & 10.618 & 0.59 \\
\hline
\multicolumn{4}{c}{CsInF$_{3}$} \\
\hline
$C2/m$-$a$ (No. 12)     & -115.011 & 0     & 0.78 \\
$I4/m$ (No. 87)         & -114.999 & 0.012 & 0.78 \\
$R\bar{3}$ (No. 148)    & -114.995 & 0.016 & 0.78 \\
$P\bar{1}$-$a$ (No. 2)  & -114.983 & 0.028 & 0.78 \\
$Fm\bar{3}m$ (No. 225)  & -114.228 & 0.783 & 0.77 \\
\end{tabular}
\label{stability-F}
\end{ruledtabular}
\end{table}

\subsection{Thermodynamic stability}
\label{sec:results_stability}

We construct the ternary phase diagrams for AInF$_{3}$ compounds
as shown in Fig.~\ref{ternary-F}.
All lowest-energy AInF$_{3}$ compounds
are thermodynamically stable (Table~\ref{stability-F}).
The determinant reactions for AInF$_{3}$ are
\begin{eqnarray}
\text{LiF} + \text{In} + 2~\text{LiInF}_{4} &\rightarrow& 3~\text{LiInF}_{3},
\nonumber \\
3~\text{NaF} + \text{In} + 2~\text{InF}_{3} &\rightarrow& 3~\text{NaInF}_{3},
\nonumber \\
4~\text{K}_{3}\text{InF}_{6} + 3~\text{KIn}_{2}\text{F}_{7} + 5~\text{In} &\rightarrow& 15~\text{KInF}_{3},
\nonumber \\
5~\text{RbF} + 3~\text{In} + 2~\text{Rb}_{2}\text{In}_{3}\text{F}_{11}
&\rightarrow& 9~\text{RbInF}_{3},
\nonumber \\
3~\text{CsF} + \text{In} + 2~\text{InF}_{3} &\rightarrow& 3~\text{CsInF}_{3}.
\nonumber
\end{eqnarray}
Given the above determinant reactions,
we can estimate the energy above/below hull ($E_{\text{Hull}}$) for AInF$_{3}$
and the results are shown in Table~\ref{stability-F}.
Since all the lowest-energy AInF$_{3}$ compounds
have substantial energies below hull, they are most likely to form.

\begin{figure*}[t]
\includegraphics[width=15.3 cm]{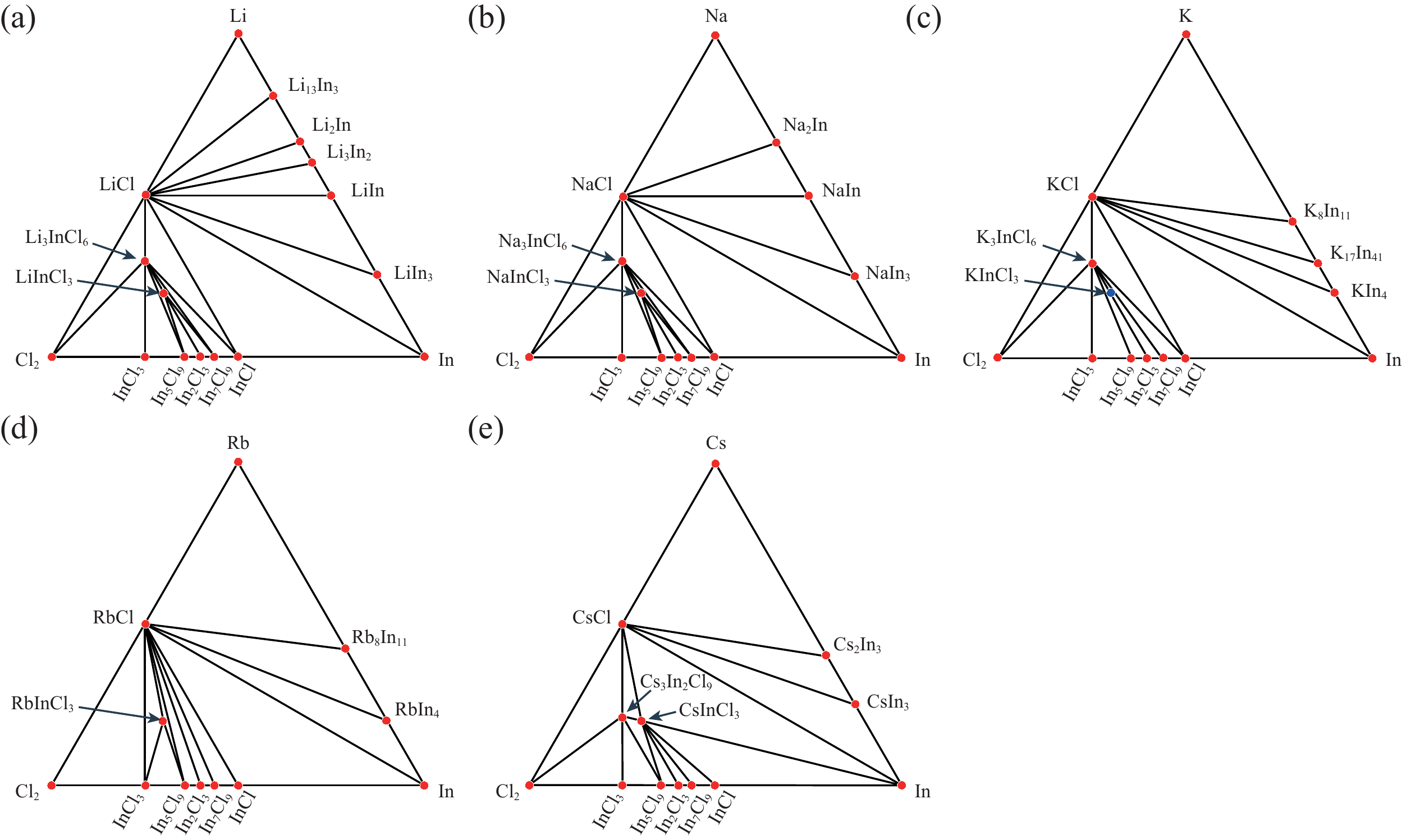}
\caption{(Color Online)
Ternary phase diagrams for (a) LiInCl$_{3}$,
(b) NaInCl$_{3}$, (c) KInCl$_{3}$, (d) RbInCl$_{3}$,
and (d) CsInF$_{3}$.
Red and blue filled circles represent
thermodynamically stable and unstable phases, respectively.
KInCl$_{3}$ is thermodynamically unstable
with a quite small energy above hull of 0.617 meV/atom,
which is marginally stable.
}
\label{ternary-Cl}
\end{figure*}

We are now in a position to discuss
the possibility of synthesis of suggested AInF$_{3}$ materials.
To do that,
we made a probabilistic model to estimate
the existence probability of a material \cite{Adler17}.
If a material has a significant energy below hull,
the existence probability becomes larger,
indicating synthesis of the material would be likely possible.
LiInF$_{3}$, NaInF$_{3}$, KInF$_{3}$, and RbInF$_{3}$
have energy below hull of -30, -54, -21, and -50 meV/atom, respectively,
which correspond to the existence probability of 0.56, 0.64, 0.52, and 0.62, respectively.
Since these existence probabilities are large enough,
these AInF$_{3}$ compounds are expected to be synthesized in laboratory.
Especially CsInF$_{3}$ phases have energies below hull of order of -100 meV/atom,
which corresponds to the existence probability of 0.78.
Since five candidate structures have small energy differences among them
(within 1 meV/atom),
existence probabilities for the structures are almost same as 0.78
(see Table.~\ref{stability-F}).
As mentioned before, the synthesis of the crystal structure strongly depends on
the synthesis method and condition.
In the case of CsInF$_{3}$,
one of the five candidate structures would be chosen and synthesized
depending on the synthesis condition.
Otherwise, the material exists in several different polymorphs.

\begin{table}[b]
\caption{
Same as Table.~\ref{stability-F} for AInCl$_{3}$ compounds.
}
\begin{ruledtabular}
\begin{tabular}{c|c|c|c}
Space group & $E_{\text{Hull}}$ (meV) & $\Delta E$ (meV) & Prob. \\
\hline
\multicolumn{4}{c}{LiInCl$_{3}$} \\
\hline
$C2$ (No. 5)           & -23.532 & 0     & 0.53 \\
$C2/m$-$c$ (No. 12)    & -23.228 & 0.304 & 0.53 \\
$R\bar{3}$ (No. 148)   & -21.669 & 1.863 & 0.52 \\
$P\bar{1}$-$a$ (No. 2) & -21.648 & 1.885 & 0.52 \\
\hline
\multicolumn{4}{c}{NaInCl$_{3}$} \\
\hline
$C2$ (No. 5)           & -4.727 & 0     & 0.45 \\
$C2/m$-$c$ (No. 12)    & -3.410 & 1.316 & 0.45 \\
$R\bar{3}$ (No. 148)   & 0.114  & 4.841 & 0.44 \\
$P\bar{1}$-$a$ (No. 2) & 0.130  & 4.857 & 0.44 \\
\hline
\multicolumn{4}{c}{KInCl$_{3}$} \\
\hline
$C2/m$-$b$ (No. 12)     & 0.617 & 0     & 0.43 \\
$P\bar{1}$-$b$ (No. 2)  & 8.450 & 7.833 & 0.41 \\
\hline
\multicolumn{4}{c}{RbInCl$_{3}$} \\
\hline
$P\bar{1}$-$b$ (No. 2)  & -79.011 & 0     & 0.70 \\
$C2/m$-$b$ (No. 12)     & -74.430 & 4.581 & 0.69 \\
\hline
\multicolumn{4}{c}{CsInCl$_{3}$} \\
\hline
$P\bar{1}$-$b$ (No. 2)  & -20.112 & 0     & 0.52 \\
$C2/m$-$b$ (No. 12)     & -17.336 & 2.776 & 0.51 \\
$I4/m$ (No. 87)         & -11.657 & 8.455 & 0.48 \\
\end{tabular}
\label{stability-Cl}
\end{ruledtabular}
\end{table}

Figure~\ref{ternary-Cl} shows the ternary phase diagrams for AInCl$_{3}$ compounds.
The determinant reactions for AInCl$_{3}$ are
\begin{eqnarray}
\text{Li}_{3}\text{InCl}_{6} + \text{In}_{2}\text{Cl}_{3} &\rightarrow& 3~\text{LiInCl}_{3},
\nonumber \\
\text{Na}_{3}\text{InCl}_{6} + \text{In}_{2}\text{Cl}_{3} &\rightarrow& 3~\text{NaInCl}_{3},
\nonumber \\
\text{K}_{3}\text{InCl}_{6} + \text{In}_{2}\text{Cl}_{3} &\rightarrow& 3~\text{KInCl}_{3},
\nonumber \\
6~\text{RbCl} + \text{InCl}_{3} + \text{In}_{5}\text{Cl}_{9} &\rightarrow& 6~\text{RbInCl}_{3},
\nonumber \\
\text{Cs}_{3}\text{In}_{2}\text{Cl}_{9} + \text{In} &\rightarrow& 3~\text{CsInCl}_{3}.
\nonumber
\end{eqnarray}
Given the above determinant reactions,
we can estimate the energy above/below hull ($E_{\text{Hull}}$) for AInCl$_{3}$
and the results are shown in Table~\ref{stability-Cl}.
All the lowest-energy AInCl$_{3}$ compounds are thermodynamically stable
except for KInCl$_{3}$.
KInCl$_{3}$ has energy above hull of 0.617 meV/atom
and the corresponding existence probability is 0.43,
which is quite significant.
Therefore, KInCl$_{3}$ is marginally stable
and is still expected to be synthesized in laboratory.
The other AInCl$_{3}$ compounds
have energy below hull of -23, -4, -79, and -20 meV/atom
for Li, Na, Rb, and Cs compounds,
which correspond to the existence probability of 0.53, 0.45, 0.70, and 0.52, respectively.
Since these existence probabilities are large enough,
these AInCl$_{3}$ compounds are expected to be synthesized in laboratory.
Both LiInCl$_{3}$ and NaInCl$_{3}$ have tiny energy differences
(within 5 meV/atom) among four candidate space groups,
$C2$, $C2/m$-$c$, $R\bar{3}$, and $P\bar{1}$-$a$.
For RbInCl$_{3}$ and CsInCl$_{3}$,
two candidate space groups, $P\bar{1}$-$b$ and $C2/m$-$b$,
also have tiny energy differences (within 5 meV/atom) between them.
Therefore AInCl$_{3}$ (A = Li, Na, Rb, Cs)
could be synthesized into one of the candidate structures
depending on the synthesis condition.
Otherwise, the material exists in several different polymorphs.
For CsInCl$_{3}$,
the third lowest-energy structure ($I4/m$) has a small energy difference
of $\sim$8 meV with the lowest one and still has a possibility to form
due to the considerable existence probability of 0.48
as shown in Table.~\ref{stability-Cl}.

\begin{figure}[b]
\includegraphics[width=8.0 cm]{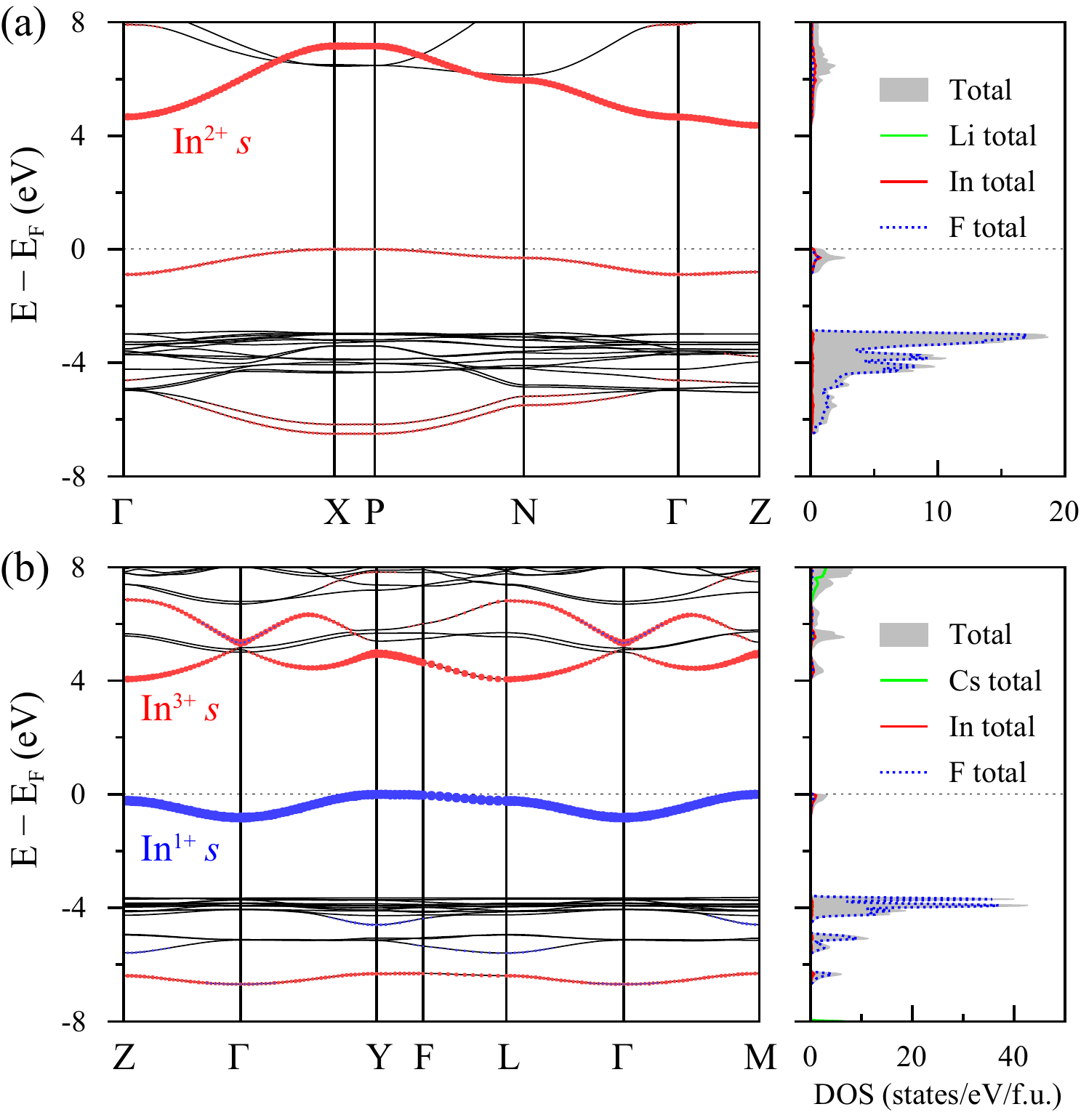}
\caption{(Color Online)
Electronic structures of (a) $I4/mmm$-$a$ LiInF$_{3}$
and (b) $C2/m$-$a$ CsInF$_{3}$ calculated by the mBJ method.
In$^{2+}$ $s$-orbital character
is marked in red in (a) and
In$^{3+}$ and In$^{1+}$ $s$-orbital characters
are marked in red and blue in (b), respectively,
in the band structures.
}
\label{band}
\end{figure}

\begin{figure}[t]
\includegraphics[width=8.0 cm]{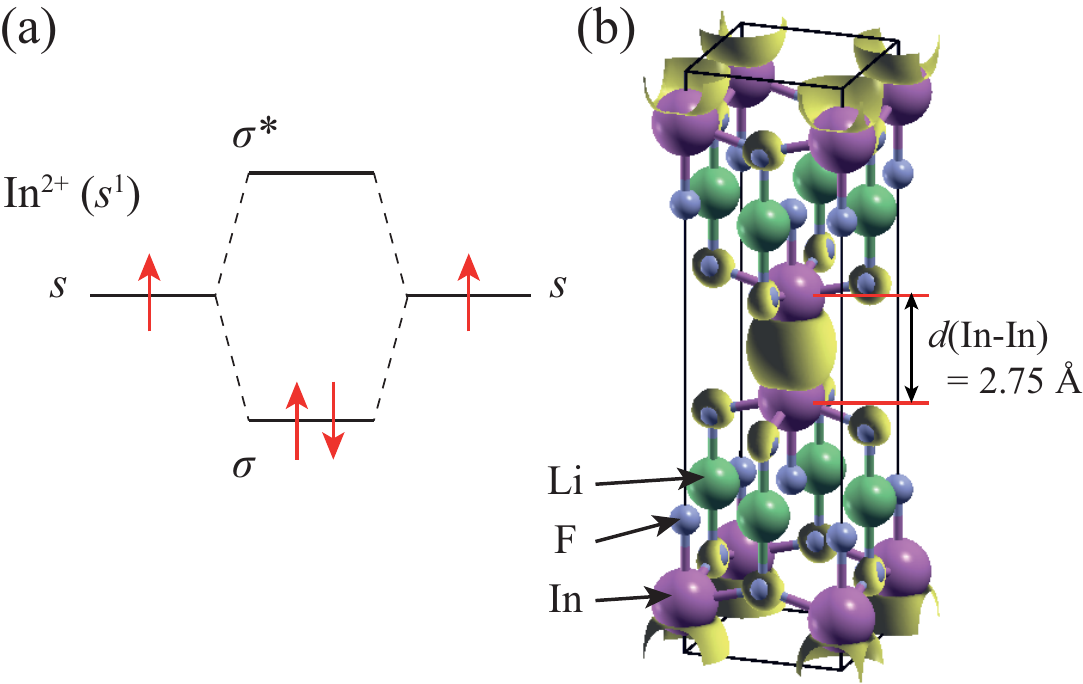}
\caption{(Color Online)
Insulating mechanism of $I4/mmm$-LiInF$_{3}$.
(a) Systematic band diagram for a band gap opening.
Two $s$ orbitals from two adjacent In$^{2+}$ atoms
form $ss\sigma$-bonding to open the band gap.
(b) Electronic charge density corresponding to the $ss\sigma$-bonding.
The distance between two adjacent In$^{2+}$ atoms
is 2.75 {\AA}.
}
\label{charge}
\end{figure}

\subsection{Electronic structure and superconductivity}
\label{sec:results_electronics}

We have calculated the electronic structures of AInF$_{3}$ compounds
and found several similarities and differences among them.
In Fig.~\ref{band}, we present two representative electronic structures of
$I4/mmm$-$a$ LiInF$_{3}$ and $C2/m$-$a$ CsInF$_{3}$.
The similarities and differences in the electronic structures of AInF$_{3}$
will be discussed later.

Figure~\ref{band} shows that
both LiInF$_{3}$ and CsInF$_{3}$ have an insulating phase
with a indirect band gap of 4.37 and 4.06 eV, respectively, based on the mBJ method
(GGA results give 2.30 and 3.05 eV, respectively).
In addition, In $s$ and F $p$ orbitals are well separated in the energy window,
which leads to only small hybridization between In $s$ and F $p$ orbitals.
The only notable hybridization is between In $s$ and F $p_{\sigma}$ orbitals,
which produces the low-lying bonding state located at -8 $\sim$ -6 eV
as shown in Fig.~\ref{band}.
However, the low-lying bonding state has dominant F $p_{\sigma}$ character
and quite small In $s$ character (see the density of states in Fig.~\ref{band}).
Due to the quite small hybridization between In $s$ and F $p$ orbitals,
we can mention that LiInF$_{3}$ and CsInF$_{3}$ have
In$^{2+}$ ($5s^1$) and multivalent In$^{1+}$ ($5s^2$) and In$^{3+}$ ($5s^0$) configurations, respectively.
We would like to note that charge disproportionation in BaBiO$_{3}$ \cite{Cox76,Cox79}
($2\text{Bi}^{4+} \rightarrow \text{Bi}^{3+} + \text{Bi}^{5+}$) was questioned
due to the significant hybridization between Bi $6s$ and O $2p$ orbitals.
The alternative scenario of oxygen hole pairs condensation
($2\text{Bi}^{3+}\underbar{L} \rightarrow \text{Bi}^{3+}\underbar{L}^2 + \text{Bi}^{3+}$, where $\underbar{L}$ represents a ligand hole) was proposed \cite{Sawatzky15,Sawatzky18,Dalpian18}.
In the case of CsInF$_{3}$, however,
hybridization between In $s$ and F $p$ orbitals is quite small.
Hence, the ligand hole pairs condensation scenario is safely ruled out and
valence/charge disproportionation is clearly realized.

The electronic structures of other low-energy AInF$_{3}$ compounds including
$I4/mmm$-$a$ NaInF$_{3}$, $I4/mmm$-$b$ KInF$_{3}$, and
$I4/mmm$-$b$ RbInF$_{3}$ (not shown) are quite
similar to that of $I4/mmm$-$a$ LiInF$_{3}$
except for the different size of the band gap.
The sizes of the indirect band gap for Na, K, and Rb compounds
obtained from the mBJ method are
3.89, 3.91, and 3.67 eV,
(GGA results give 1.77, 1.90, and 1.90 eV), respectively,
which are smaller than that for LiInF$_{3}$.
For CsInF$_{3}$, the five candidate structures,
which are $C2/m$-$a$, $I4/m$, $R\bar{3}$, $P\bar{1}$-$a$, and $Fm\bar{3}m$,
also have similar electronic structure in the following manner:
(i) In $s$ and F $p$ orbitals are well separated in the energy window
(In $s$ and F $p$ orbitals lie -1 $\sim$ 7 eV and -7 $\sim$ -4 eV, respectively),
(ii) quite small hybridization between In $s$ and F $p$ orbitals is realized,
(iii) multivalent In$^{1+}$ ($5s^2$) and In$^{3+}$ ($5s^0$) valence skip configuration is clearly shown,
and (iv) the five candidate structures have similar band gaps of $\sim$4.0 eV
(GGA results give almost same band gaps of $\sim$3.0 eV).

Both $I4/mmm$-$a$ LiInF$_{3}$ and $C2/m$-$a$ CsInF$_{3}$ show an insulating behavior,
however, the insulating mechanism is different
(The insulating mechanism of the Na, K, and Rb compounds
is essentially same as the Li compound).
As shown in Fig.~\ref{band}(b),
$C2/m$-$a$ CsInF$_{3}$ exhibits the mix-valent In$^{1+}$ ($5s^{2}$) and
In$^{3+}$ ($5s^{0}$) character \cite{chdiff},
where In$^{1+}$ has a completely filled $s$-band
and In$^{3+}$ has an (nearly) empty $s$-band
(there is a small amount of In$^{3+}$ $s$-orbital character in the occupied band
due to the small hybridization between In $s$ and F $p_{\sigma}$ orbitals).
Therefore, the insulating mechanism in $C2/m$-$a$ CsInF$_{3}$ is straightforward:
the mix-valent character makes the insulating phase.
For the case of $I4/mmm$-$a$ LiInF$_{3}$,
In$^{2+}$ ($5s^{1}$) character is clearly shown in Fig.~\ref{band}(a),
where In$^{2+}$ has the half-filled $s$-band,
which is not straightforward to understand the band gap formation.
In order to investigate the insulating mechanism
in $I4/mmm$-$a$ LiInF$_{3}$ thoroughly,
we plot the charge density with the energy range between -2 and 0 eV
(the Fermi level is set to be zero) in Fig.~\ref{charge}(b),
which corresponds to the $ss\sigma$-bonding character.
Therefore, two adjacent In atoms are dimerized
with the bond length of 2.75 {\AA}
to produce $ss\sigma$ (bonding) and $ss\sigma^{*}$ (anti-bonding) states,
and each $s$-electron in In$^{2+}$ completely fills the $ss\sigma$-bonding state (Fig.~\ref{charge}(a)).
The band gap size in $I4/mmm$-$a$ LiInF$_{3}$ indicates
the splitting between $ss\sigma$ (bonding) and $ss\sigma^{*}$ (anti-bonding) states.

The electronic structures of all AInCl$_{3}$ compounds
show insulating behavior as well (not shown).
AInCl$_{3}$ could be classified into two groups:
LiInCl$_{3}$ and NaInCl$_{3}$ are in the same group and
KInCl$_{3}$, RbInCl$_{3}$, and CsInCl$_{3}$ belong to the other group.
For each group, they share the same low-energy candidate structures:
both LiInCl$_{3}$ and NaInCl$_{3}$ have four candidate structures
with space groups of $C2$, $C2/m$-$c$, $R\bar{3}$, and $P\bar{1}$-$a$,
and KInCl$_{3}$, RbInCl$_{3}$, and CsInCl$_{3}$ have two candidate structures
with space groups of $P\bar{1}$-$b$ and $C2/m$-$b$.
Not only the electronic structures but also the size of the band gap
are almost the same among the candidate structures for each AInCl$_{3}$ compound.
The band gaps obtained from the mBJ method are
3.5, 3.5, 4.2, 4.3, and 4.3 eV for Li, Na, K, Rb, and Cs compounds, respectively
(GGA gives the band gaps of 2.8, 2.7, 2.9, 3.0, and 3.0 eV, respectively).
For both LiInCl$_{3}$ and NaInCl$_{3}$,
their four candidate structures possess two different sizes of InCl$_{6}$ octahedra
(bond disproportionation)
and show the multivalent In$^{1+}$ ($5s^2$) and In$^{3+}$ ($5s^0$)
valence skip configuration.
Note that In $s$ and Cl $p$ orbitals are well separated in the energy window
and only a small hybridization between them is realized.
Since the Cl $p$ orbital has higher energy than the F $p$ orbital,
the hybridization between In $s$ and Cl $p_{\sigma}$ orbitals is somewhat bigger than
that between In $s$ and F $p_{\sigma}$ orbitals.
However, the hybridization between In $s$ and Cl $p$ is still quite small,
hence the ligand hole pairs condensation scenario is safely ruled out
and valence/charge disproportionation is clearly realized.
Interestingly, two $s$ bands originating from In$^{1+}$ and In$^{3+}$ atoms
are almost flat (their bandwidths are smaller than 1 eV)
and are located at near the Fermi level and $\sim$4 eV, respectively.
For KInF$_{3}$, RbInF$_{3}$, and CsInF$_{3}$ compounds,
their candidate structures with space groups of $P\bar{1}$-$b$ and $C2/m$-$b$
have symmetrically equivalent In atoms in the unit cell.
Therefore, valence/charge disproportionation is not possible
in their candidate structures.
Alkali metal (K, Rb, Cs) and Cl atoms
strongly prefer the valence states of 1+ and 1-, respectively,
leading to In atoms having 2+ valence state.
The insulating mechanism of these compounds is essentially the same as
the one discussed in LiInF$_{3}$,
and the band gap is formed between
the fully occupied $ss\sigma$ bonding and totally empty $ss\sigma^*$
anti-bonding states.

\begin{figure}[t]
\includegraphics[width=8.0 cm]{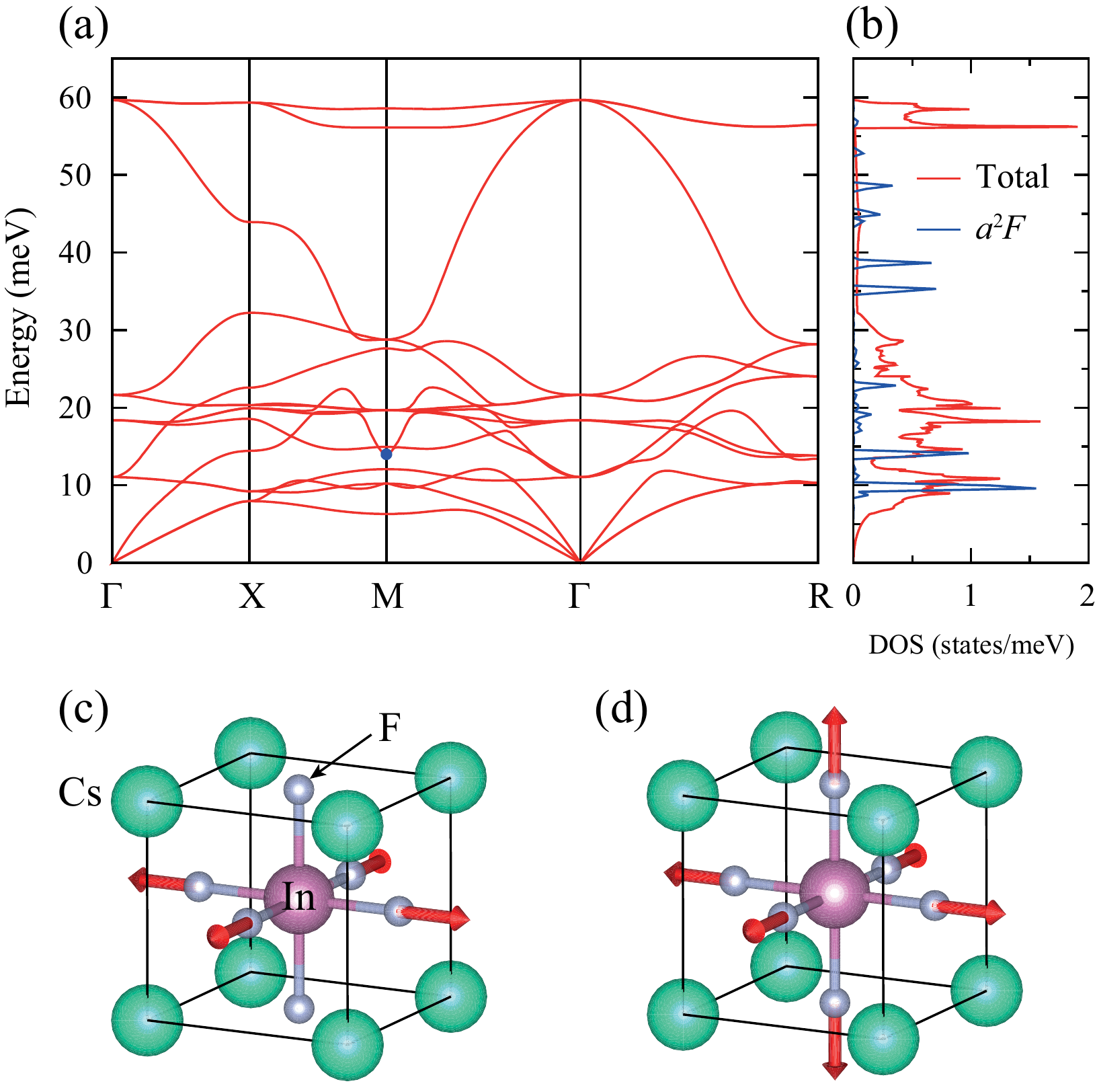}
\caption{(Color Online)
(a) Phonon dispersion and (b) the corresponding phonon density of state
and the Eliashberg function $a^2F(\omega)$
of 0.5 hole-doped cubic $Pm\bar{3}m$-CsInF$_{3}$
under pressure $P$ = 32 GPa ($a$ = 4.233 {\AA}).
The unit cell of cubic $Pm\bar{3}m$-CsInF$_{3}$ is shown in (c) and (d).
The arrows in (c) and (b) show phonon normal modes at $q = M$ and $R$, respectively,
having a significant electron-phonon coupling constant.
The phonon normal mode having
the largest electron-phonon coupling strength $\lambda_{q\nu}$
is the F stretching phonon mode shown in (c) and the size of $\lambda_{q\nu}$
is marked by a blue circle in (a).
}
\label{eph}
\end{figure}

As discussed earlier, all AInX$_{3}$ (A = alkali metals, X = F or Cl) compounds
have insulating behavior with a significant band gap of approximately 4 eV
(based on the mBJ method).
Hence, these compounds are not suitable for photovoltaic materials
(appropriate band gap size is ranged from 1 to 2 eV)
without any band gap tuning like, for example, applying strain or pressure.

Now, we discuss the possible superconductivity in these new In compounds.
As shown in Fig.~\ref{band},
hole doping gives the higher electronic density of states
than electron doping in both LiInF$_{3}$ and CsInF$_{3}$.
Therefore, hole doping would be expected to give
higher superconducting transition temperature $T_{c}$ than electron doping.

We calculated the electron-phonon coupling constant $\lambda$
in 10 \% hole-doped $I4/mmm$-$a$ LiInF$_{3}$.
We found that $\lambda$ = 0.324
and the logarithmic average phonon frequency $\omega_{log}$ = 304 K,
which gives the superconducting transition temperature $T_{c}$ = 0.293 (0.02) K
when we use the effective Coulomb repulsion parameter $\mu^{*}$ = 0.10 (0.15)
in the McMillan-Allen-Dynes formula \cite{Allen75}.

The emergence of superconductivity was reported in
charged doped cubic perovskite BaBiO$_{3}$ \cite{Yin13}
and CsTiX$_{3}$ (X = F or Cl) \cite{Yin13epl} materials.
Therefore, the simple cubic perovskite CsInF$_{3}$ ($Pm\bar{3}m$; No. 221)
is also tested for possible superconductivity.
At ambient pressure, it is mechanically unstable with imaginary phonon softening
at $X$ and $R$ points (not shown) even with sufficient doping.
The unstable phonon mode at $R$ point
is the InF$_{6}$ octahedron breathing mode as shown in Fig.~\ref{eph}(d).
With sufficient doping and at high pressure,
the structure distortions and valence disproportionation
realized in CsInF$_{3}$ (the second row of Fig.~\ref{crystal}) could be suppressed
and CsInF$_{3}$ could crystallize in the simple cubic perovskite structure
accompanied by a transition to the metallic phase.
This bears a close resemblance to the phase diagram of BaBiO$_{3}$ \cite{Cox76,Cox79}.

\begin{table}[t]
\caption{
Phonon related physical parameters for 0.5 hole-doped simple cubic CsInF$_{3}$
under pressure $P$ = 32 GPa.
The reduced electron-phonon matrix element (REPME),
the total electron-phonon coupling $\lambda$,
the average phonon frequency $\omega_{\text{log}}$,
and the superconducting transition temperature $T_{c}$
are evaluated in both GGA and HSE06 methods.
The REPME is obtained for the F stretching mode shown in Fig.~\ref{eph}(c),
which is the most important vibration mode.
The temperature $T_{c}$ is estimated
by using the McMillan-Allen-Dynes formula \cite{Allen75}
with $\mu^{*}$ = 0.10.
$T_{c}$ with $\mu^{*} = 0.15$ is also provided in the parentheses.
Note that the temperature $T_{c}$ is not sensitive to $\mu^{*}$ within the confidence interval ranged from 0.10 to 0.15.
}
\begin{ruledtabular}
\begin{tabular}{c|c|c}
& GGA & HSE06 \\
\hline
REPME (eV/{\AA}) & 5.00 & 6.95 \\
$\lambda$ & 1.80 & 3.48 \\
$\omega_{\text{log}}$ (K) & 157 & 124 \\
$T_{c}$ (K) & 21 (18) & 24 (22)\\
\end{tabular}
\label{eph-HSE}
\end{ruledtabular}
\end{table}

Figure \ref{eph}(a) shows the phonon dispersion of 0.5 hole-doped cubic perovskite CsInF$_{3}$ under pressure of $P$ = 32 GPa
(hole-doped CsInF$_{3}$ is possible
through a certain amount of oxygen substitution for fluorine, for example, CsIn(F$_{1-x}$O$_{x}$)$_{3}$, or Cs vacancy could introduce hole doping).
The F vibrational mode has the highest energy due to its small mass.
However, at particular phonon momentum $q$ (especially at $q = M, R$),
this F vibrational mode shows quite significant phonon softening
indicating fairly large electron-phonon coupling.
The phonon momentum $q$ and mode $\nu$ dependent
electron-phonon coupling strength $\lambda_{q\nu}$ are
7.24 and 0.19 for the soft phonon modes
shown in Figs.~\ref{eph}(c) and (d), respectively.
The (total) electron-phonon coupling constant $\lambda$ is 1.80
and the logarithmic average phonon frequency $\omega_{log}$ = 157 K.
The McMillan-Allen-Dynes formula \cite{Allen75} with $\mu^{*}$ = 0.10 (0.15)
gives $T_{c}$ = 21 (18) K.

Conventional electronic descriptions based on LDA/GGA
underestimates the electron-phonon coupling constant $\lambda$
in Ba$_{1-x}$K$_{x}$BiO$_{3}$ and fail to explain its
high-temperature superconductivity \cite{Meregalli98}.
Yin \emph{et al.} found that static correlation
(which is captured within GW or the hybrid HSE06 functional)
enhances the electron-phonon coupling constant $\lambda$
and describes the transition temperature $T_{c}$ properly \cite{Yin13}.
Motivated by the above study \cite{Yin13},
the effect of static correlation on superconductivity in CsInF$_{3}$ is also tested.
Since the F stretching phonon mode (Fig.~\ref{eph}(c)) is the most important
to account for superconductivity,
we calculated a reduced electron-phonon matrix element (REPME) \cite{Yin13}
for the phonon mode.
Table~\ref{eph-HSE} shows that
static correlation enhances the REPME by a factor of $\sim$1.4
and gives rise to larger electron-phonon coupling constant $\lambda$ = 3.48
and $T_{c} \approx$ 24 K.
These results encourage the pursuit of experimental synthesis.

\section{Summary and conclusions}
\label{sec:conclusion}

Using computational methods we identified
nine new In compounds AInX$_{3}$ (A = alkali metals, X = F or Cl)
and investigated their   crystal structures  and physical properties.
Two distinct insulating mechanisms are realized in the new In compounds.
For the compounds that have different sizes of InX$_{6}$ octahedra
(bond disproportionation),
which are CsInF$_{3}$, LiInCl$_{3}$, and NaInCl$_{6}$,
they show multivalent In$^{3+}$ ($5s^0$) and In$^{1+}$ ($5s^2$)
valence skip configuration.
The insulating phase is induced from bond and valence disproportionation.
The other In compounds,
AInF$_{3}$ (A = Li, Na, K, Rb) and AInCl$_{3}$ (A = K, Rb, Cs),
have symmetrically equivalent In sites in their unit cell
and hence both bond and valence disproportionation are not possible.
These In compounds have the divalent valence state of In,
which is very rare but thermodynamically stable.
In these compounds, two adjacent In atoms are dimerized
to produce $ss\sigma$ bonding and $ss\sigma^*$ antibonding states.
Each $s$-electron in In$^{2+}$ completely fills the $ss\sigma$ bonding states.
Therefore, the band gap sizes in these compounds indicate
the splitting between $ss\sigma$ bonding and $ss\sigma^*$ antibonding states.

These compounds are a new arena for investigating
charge, valence, and bond disproportionation.
They are similar to the CsTlX$_{3}$ (X = F or Cl),
which had also been predicted theoretically and synthesized experimentally
\cite{Yin13epl,Retuerto13},
but do not have the toxicity problems associated with Tl,
and therefore can be synthesized more easily.
All these AInX$_{3}$ compounds are insulators
with a sizable indirect band gap of the order of $\sim$4 eV,
which makes them unsuitable for photovoltaic materials.
We also studied the possible superconductivity in these new In compounds.
With sufficient doping and at high pressure,
CsInF$_{3}$ crystallizes in the simple cubic perovskite structure
and has a significant superconducting transition temperature of $T_{c} \approx$ 24 K.

\begin{figure}[t]
\includegraphics[width=7.0 cm]{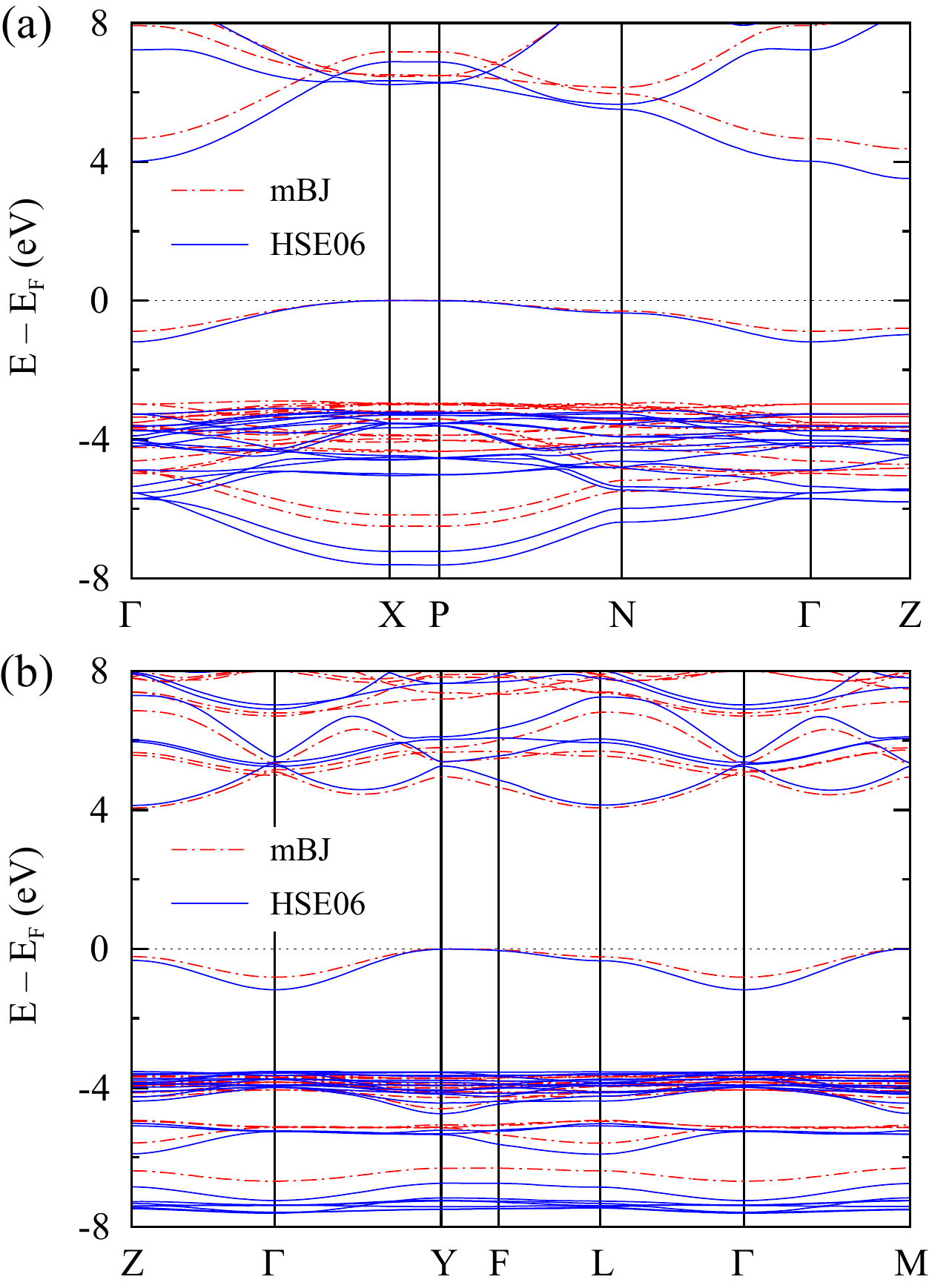}
\caption{(Color Online)
Comparison between mBJ and HSE06 methods
for (a) $I4/mmm$-$a$ LiInF$_{3}$ and (b) $C2/m$-$a$ CsInF$_{3}$.
Both mBJ and HSE06 methods give the similar band gap.
}
\label{hse06}
\end{figure}

\begin{acknowledgments}
We thank Ran Adler, Ian Fisher, Martha Greenblatt, and Xiaoyan Tan for fruitful discussions.
This work was supported  by the
Air Force Office of Scientific Research supported MURI under Grant No. FA9550-14-1-0331.
\end{acknowledgments}

\section{Appendix}
We have estimated the band gaps of the semiconducting systems within the mBJ method.
In Appendix, we discuss the computational validity of the mBJ method.
For the semiconducting systems,
the mBJ method is efficient to improve the band gap but is empirical.
On the other hand, the rather computationally expensive HSE06 method
is well physically constructed and is more accurate on this aspect.
Here, we compute the electronic structure within the HSE06 method
and compare it with the mBJ method.

Figure \ref{hse06} shows a comparison between the mBJ and HSE06 methods
for two representative electronic structures of
$I4/mmm$-$a$ LiInF$_{3}$ and $C2/m$-$a$ CsInF$_{3}$.
For $I4/mmm$-$a$ LiInF$_{3}$, mBJ and HSE06 methods
give indirect band gaps of 4.37 and 3.51 eV, respectively.
In addition, both methods give indirect band gaps of 4.06 and 4.15 eV,
respectively, for $C2/m$-$a$ CsInF$_{3}$.
Since both mBJ and HSE06 methods give a similar band gap
for the semiconducting systems investigated in this study,
the band gaps estimated from the mBJ method are reliable.


\begin{thebibliography}{99}
\bibitem{structure-set}
    We have performed USPEX simulations for several AInX$_{3}$ materials
    including LiInF$_{3}$, RbInF$_{3}$, CsInF$_{3}$, LiInCl$_{3}$, RbInCl$_{3}$,
    and CsInCl$_{3}$.
    After the USPEX simulations, we have collected the crystal structures
    with energy differences from the lowest energy within 10 meV/atom
    for each material
    and made a structure set as shown in Fig.~\ref{crystal}.
\bibitem{chdiff}
    The charge density difference between In$^{1+}$ and In$^{3+}$ atoms
    in $C2/m$-CsInF$_{3}$ is small ($\sim$0.21 $e$)
    in the DFT calculations.
    Note that there is no real difference in $d$ occupations
    even in well-known charge-order transition metal oxides.
    See Ref.~[\onlinecite{Quan12}] in more detail.
\end{thebibliography}
\end{document}